\documentstyle[epsfig,aps,prl,multicol]{revtex}
\begin{document}

\title{Classical mechanics technique for quantum linear response}

\author{E.V. Tsiper}

\address{Department of Physics, SUNY at Stony Brook, Stony Brook, NY
11794.}

\date{March 9, 2000}

\maketitle

 \begin{abstract}
 It is shown that the lowest excitation energies of a quantum
many-fermion system in the random phase approximation (RPA) can be
obtained by minimizing an effective classical energy functional.  The
minimum can be found very efficiently using generalized Lanczos
technique.  Application of the new technique to molecular spectra
allows to compute excited states at the expense comparable to the
ground-state calculations.  As an example, the first-principle RPA
excitation spectrum of C60 molecule is computed taking into account
all 240 valence electrons in the full valence space of the molecule.
The results match linear absorption experiment within percents.
 \end{abstract}

\begin{multicols}{2}

Random phase approximation is central to the theory of electronic
excitations in molecules and in extended systems \cite{blaizot}.  More
generally, it resides in the core of the theory of linear response of
correlated many-particle systems, and is widely used to describe
correlation effects in the excitations of nuclei
\cite{thouless,tohyama}, molecules \cite{zerner,muk_sci},
semiconductor quantum wells \cite{wang}, quantum dots
\cite{berkovits}, and bulk materials \cite{karlsson}.  Large systems
of RPA-type equations are of particular significance in photochemistry
of large molecules.  Understanding such processes as photosynthesis
and light reception in vision requires detailed description of the
evolution of large biological molecular complexes upon optical
excitation \cite{rhodopsin}, which is governed by the configuration of
excited-state adiabatic surfaces \cite{ours_jcp}.  Yet, unlike
ground-state molecular calculations, which are now considered a
routine job with the powerful program packages at hand
\cite{gaussian}, the modeling of the excited states is a much more
difficult task \cite{large_ex}.

The reason behind this difficulty is that {\em electronic
correlations} are usually much more pronounced in the excited states.
In other words, when the ground state electronic wavefunction can be
reliably approximated using Hartree-Fock (HF) or density functional
theory (DFT), the electronic wavefunction in the excited state cannot
be described as a single Slater determinant.

RPA is one of the standard tools to treat electronic correlations in
the excited sates of quantum many-particle systems.  It belongs to a
broader family of so-called {\em time-dependent} techniques, such as
time-dependent HF (TDHF) and time-dependent DFT (TDDFT).  These
methods target directly the excitation energies of the system, by
associating these energies with the frequencies of oscillations of the
ansatz parameters when the system is driven out of equilibrium.  The
static HF or DFT ground state is the best Slater determinant that
gives minimum to a certain energy functional.  The TDHF and TDDFT
describe the time evolution of the respective Slater determinant near
the equilibrium.

An assumption is made that the time-dependent wavefunction remains a
single Slater determinant at each moment in time.  Projection of the
time-dependent Schroedinger equation onto the set of all Slater
determinants yields a set of essentially classical Hamiltonian
equations of motion \cite{blaizot}.  In linear response, when the
deviation from equilibrium is small, the motion represents small
oscillations, whose frequencies are associated naturally with the
excitation energies of the system.  In the small-oscillation limit
TDHF is equivalent to RPA.

RPA excitation energies are obtained as the eigenvalues of a
non-Hermitian matrix \cite{thouless}

\begin{equation}
\left(\begin{array}{cc}
A & B \\
-B & -A 
\end{array}\right)
\label{rpa}
\end{equation}
 The $N\times N$ symmetric matrices $A$ and $B$ describe
particle-particle interaction.  Their matrix elements are simple
combinations of the two-particle interaction matrix elements of the
Hamiltonian in the basis of HF orbitals.

Configuration interaction singles (CIS), which is another commonly
used technique to describe correlation effects, is in fact an
approximation to RPA, and can be recovered by setting $B=0$
\cite{zerner}.

In contrast to the static HF, where the number of equations scales
linearly with the number of particles, the size $2N$ of the matrix
(\ref{rpa}) grows quadratically with the size of the single-particle
Hilbert space.  This impedes diagonalization of the matrix (\ref{rpa})
for relatively large systems.  For example, RPA equations for the
singlet excited states of C60 molecule in the full valence space basis
lead to the matrix of the size $2N=28,800$.  On the other hand,
complete solution of the TDHF equations is not always necessary.  In
many cases, only a few low-energy excitonic states are of interest.

This amounts to computing only a few extremal eigenvectors of the
matrix (\ref{rpa}) --- a task similar to a standard problem in quantum
mechanics of computing a few low-energy eigenstates $\psi$ of a
Hermitian matrix $H$.  The latter problem can be solved very
efficiently using Hermitian Lanczos algorithm \cite{lanczos}, or any
of its various modifications.  In essence, the algorithm builds a
Krylov subspace ${\cal K}_n$ of the matrix $H$, and then finds the
best approximation to $\psi$ in ${\cal K}_n$ by minimizing the
expectation value of the energy $(\psi H\psi)$ \cite{parlett}.

There exist many variations of the Lanczos algorithm that allow to
find eigenpairs of non-Hermitian matrices \cite{parlett}.  These
methods, however, loose very much of the performance of the Hermitian
Lanczos algorithm, because of the lack of a {\em minimum principle}.
Indeed, no general minimum principle exists that yields eigenvalues of
non-Hermitian matrices \cite{saad}.

A lot of effort has been put into developing reliable methods for
computing selected eigenvalues of RPA-type matrices (\ref{rpa})
\cite{rettrup,jorgensen,frisch,dsma,mei,benner,narita}.  In
\cite{rettrup} the Davidson algorithm has been extended to solve RPA
equations as a general non-Hermitian eigenvalue problem.  In
\cite{jorgensen,frisch} it was modified to preserve the special paired
structure of the matrix (\ref{rpa}).  Tretiak et al.~\cite{dsma} have
developed a density-matrix-spectral-moments (DSMA) algorithm based on
generalized sum rules for the response theory.  The symplectic Lanczos
algorithm suggested by Mei\cite{mei} and improved by Benner
\cite{benner} exploits the analogy between the unitary transformations
that preserve Hermiticity and the symplectic transformations that
preserve the paired structure of (\ref{rpa}).  A Newton-Raphson-type
iterative procedure has been developed in \cite{narita}.  Finally, the
oblique Lanczos algorithm for general non-Hermitian matrices
\cite{saad} was applied to the TDHF problem in \cite{ours_L}.

It has been majorly overlooked that, although the RPA-type matrix is
non-Hermitian, its block paired structure gives it some properties
similar to the Hermitian matrices.  In particular, there does exist a
minimum principle that yields the lowest {\em positive} eigenvalue of
(\ref{rpa}).  It was suggested by Thouless back in 1961 and
reads \cite{thouless}

\begin{equation}
\omega_{\min}=
\min_{\{x,y\}}\frac{(x,y)
\left(\begin{array}{cc}
A & B \\
B & A 
\end{array}\right)
\left(\begin{array}{c}
x\\
y 
\end{array}\right)}
{|(xx)-(yy)|}
\label{th}
\end{equation}
 The minimum is to be taken over all $N$-vectors $x$ and $y$.  The
minimum always exists, since the HF stability condition keeps the
numerator positive \cite{thouless}.  Note, that the denominator can be
arbitrarily small, and therefore the expression has no maximum.

The eigenvalue equation for the matrix (\ref{rpa}) can be transformed
into the form of Hamiltonian equations of motion for classical
oscillations by substitution $T=A+B$ and $K=A-B$:

\begin{equation}
Tp=\omega q,\ \ \ \ \ Kq=\omega p,
\label{ham}
\end{equation}
 Here vectors $q$ and $p$ play the role of the conjugate canonical
coordinates and momenta, while $K$ and $T$ are the matrices of
stiffness and kinetic coefficients respectively.

The lowest frequency of a harmonic Hamiltonian system can be obtained
as a minimum of its total energy over all phase-space configurations
$\{p,q\}$ normalized by $(pq)=1$:

\begin{equation}
\omega_{\min}=\min_{(pq)=1}\frac{(pTp)}{2}+\frac{(qKq)}{2}.
\label{rr}
\end{equation}
 Indeed, variation of (\ref{rr}) with respect to $p$ and $q$ yields
Hamiltonian equations of motion (\ref{ham}).  The minimum principle
(\ref{rr}) is equivalent to the Thouless minimum principle (\ref{th}),
where $x=p+q$ and $y=p-q$.

Two terms in the right-hand side of (4) are the kinetic and the
potential energies at the configuration $\{p,q\}$.  Both are positive
for any $p$ and $q$ when the equilibrium is stable.  The positive
definiteness of $K$ and $T$ leads to the positive definiteness of the
matrix in (\ref{th}), and thus, to the HF stability condition, making
all eigenfrequencies real \cite{chi70}.

Minimum (4) can be easily found using the generalized Lanczos
recursion \cite{jetpl}

\begin{mathletters}
\begin{eqnarray}
q_{i+1}&=&\beta_{i+1}^{-1}(Tp_i-\alpha_iq_i-\beta_iq_{i-1})
\label{rrlanc1}\\
p_{i+1}&=&\delta_{i+1}^{-1}(Kq_i-\gamma_ip_i-\delta_ip_{i-1}),
\label{rrlanc2}
\end{eqnarray}
\end{mathletters}
 which generates configuration space vectors $(q_i,p_i)$ that span the
Krylov subspace of the eigenvalue problem (\ref{ham}).  When four
coefficients $\alpha_i$, $\beta_i$, $\gamma_i$, and $\delta_i$ are
chosen at each step $i$ to ensure
$(q_{i+1}p_i)=(q_{i+1}p_{i-1})=(p_{i+1}q_i)=(p_{i+1}q_{i-1})=0$, the
vectors $p_i$, $q_i$ form a biorthogonal basis, $(p_iq_j)=\delta_{ij}$
and the matrices $\widetilde K_{ij}=(q_iKq_j)$ and $\widetilde
T_{ij}=(p_iTp_j)$ are symmetric tridiagonal, with the only nonzero
matrix elements
 $\widetilde K_{ii}=\alpha_i$,
 $\widetilde K_{i,i-1}=\widetilde K_{i-1,i}=\beta_i$,
 $\widetilde T_{ii}=\gamma_i$, and
 $\widetilde T_{i,i-1}=\widetilde T_{i-1,i}=\delta_i$.
Expanding $q=\sum c_iq_i$ and $p=\sum d_ip_i$, we arrive at the
$2n\times2n$ eigenvalue problem

\begin{equation}
\widetilde Td=\widetilde\omega c,\ \ \ \ \ 
\widetilde Kc=\widetilde\omega d,
\label{rrritz}
\end{equation}
 which has the same structure as (\ref{ham}).  The lowest positive
eigenvalue $\widetilde\omega_{\min}$ of (\ref{rrritz}) give the
approximation to the true lowest frequency $\omega_{\min}$.  The
accuracy is found to improve exponentially with increasing $n$.

When the lowest-frequency normal mode $q^{(1)},p^{(1)}$ is found, the
second-lowest normal mode $q^{(2)},p^{(2)}$ can be obtained by
choosing initial vectors $q_1$ and $p_1$ orthogonal to $p^{(1)}$ and
$q^{(1)}$ respectively.  As follows from Eq.~(\ref{rr}) such a choice
causes all vectors $q_i$ and $p_i$ to remain orthogonal to $p^{(1)}$
and $q^{(1)}$.  An {\em oblique projection} can be used to correct for
the loss of orthogonality with respect to $p^{(1)}$ and $q^{(1)}$ that
may occur at large $n$.  Namely, the necessary amounts of $q^{(1)}$
and $p^{(1)}$ should be subtracted from $q_i$ and $p_i$ respectively
to ensure $(q_ip^{(1)})=(p_iq^{(1)})=0$.  Higher-frequency TDHF
solutions can be found one by one in this way.

In order to demonstrate the power of the new technique, it has been
applied to compute the excitation spectrum of the fullerene C60
molecule.  Albeit the great attention this molecule has received in
the past several years \cite{C60book}, no adequate correlated
calculation of the excited-states of C60 has been reported so far.

As noted above, CIS in the entire particle-hole configuration space
can be seen as an approximation to RPA, where the matrix $B$ in
(\ref{rpa}) is neglected.  Yet, large size of the molecle has
prevented full diagonalization of CIS matrix in the entire space.
Calculations for C60 have been reported using CIS with the CNDO/S
Hamiltonian in the truncated space of up to 1295 out of 14400
particle-hole configurations \cite{negri}, and using TDDFT with B-P86
functional \cite{kappes}.  Also a surprisingly high lowest
optically-allowed transition energy of 5.13 eV has been reported using
{\em ab-initio} Rettrup-type RPA calculation \cite{weiss}.

The technique outlined above has allowed to solve RPA equations in the
entire valence particle-hole configuration space of the molecule
(matrix size $N=2\times14400$).  INDO/S semiempirical parameterization
of the Hamiltonian \cite{indos} was used, which is essentially better
than CNDO/S, and was shown to give especially good description of the
excitation spectra of $\pi$-conjugated molecules at the CIS/RPA level
of theory \cite{zerner,muk_sci}.

 \begin{figure}
 \vskip 0.3 in
 \centerline{\epsfig{file=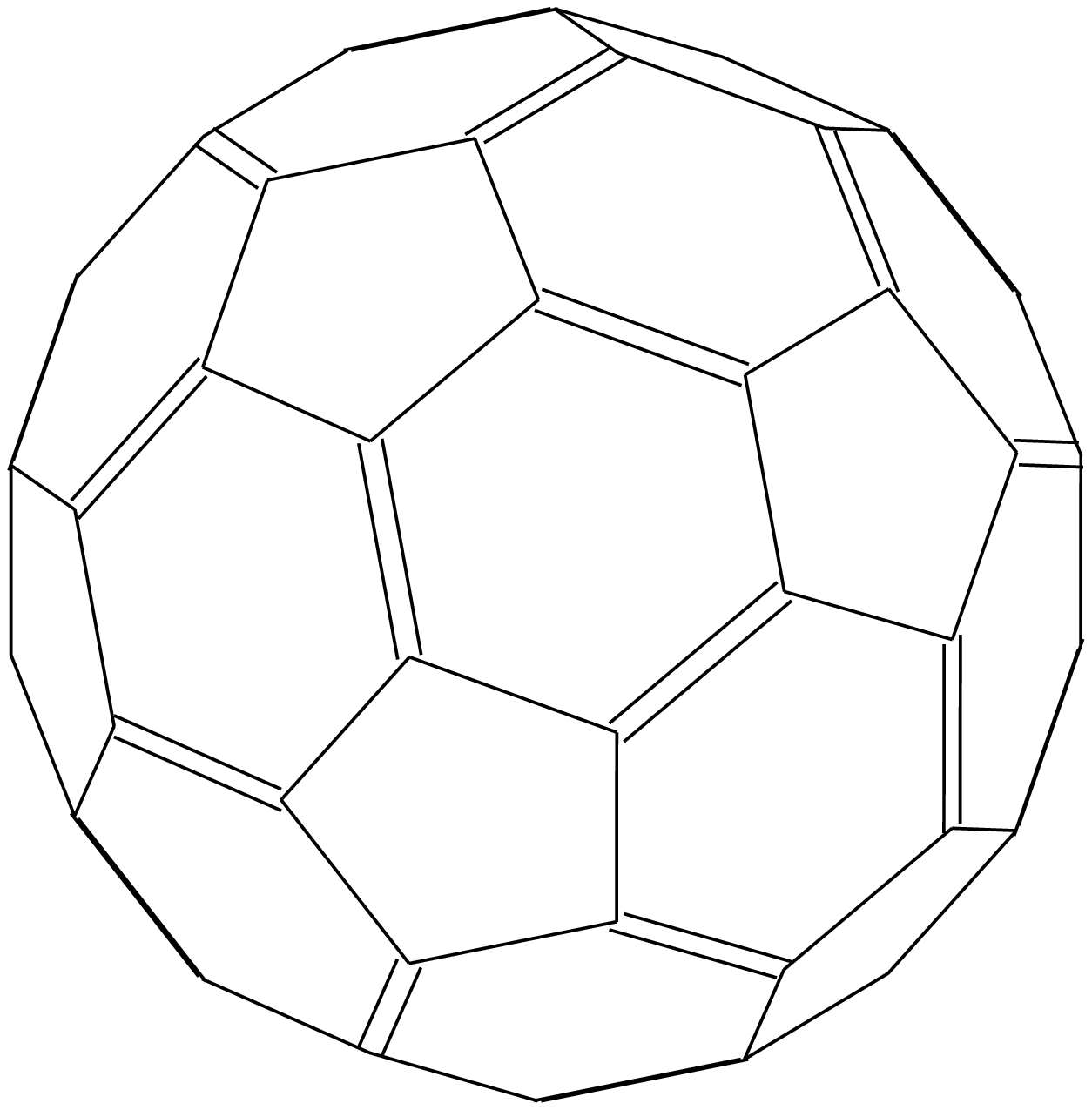,width=7cm}}
 \vskip 0.1 in
 {\small {\bf Fig.~1}. C60 geometry is completely determined by the
single and double bond lengths.}
 \end{figure}

Experimental values of 1.46 and 1.455 \AA\cite{C60book} were chosen
for single and double bond lengths respectively, which completely
determines the geometry of the molecule (see Fig.~1).  INDO/S
Hamiltonian matrix elements were generated using ZINDO program
\cite{zindo}.  Only the singlet states of C60 were studied.  The
calculation was performed on DEC Alpha 500au workstation.  Solution of
static HF equations took about 2 min.~CPU time compared to about 6
min.~per each excited state.

The present results allow for the first time for the direct comparison
to the experiment.  As shown in Table I, the energies of
optically-allowed transitions obtained are within percents from the
features observed in the linear absorption of C60 in solution
\cite{koudoumas}.  In Ref.~\onlinecite{kappes} the modes were found to
have a systematic red shift by 0.35 eV.  No systematic shift was
observed in the present study.  Almost perfect match of all transition
energies to the features seen in linear absorption allows to resolve
the controversy of the assignment of the lowest optically allowed
transition towards the value of 2.87 eV, which is opposite to the
conclusion of \cite{weiss}.

 \begin{figure}
 \vskip 0.2 in
 \centerline{\epsfig{file=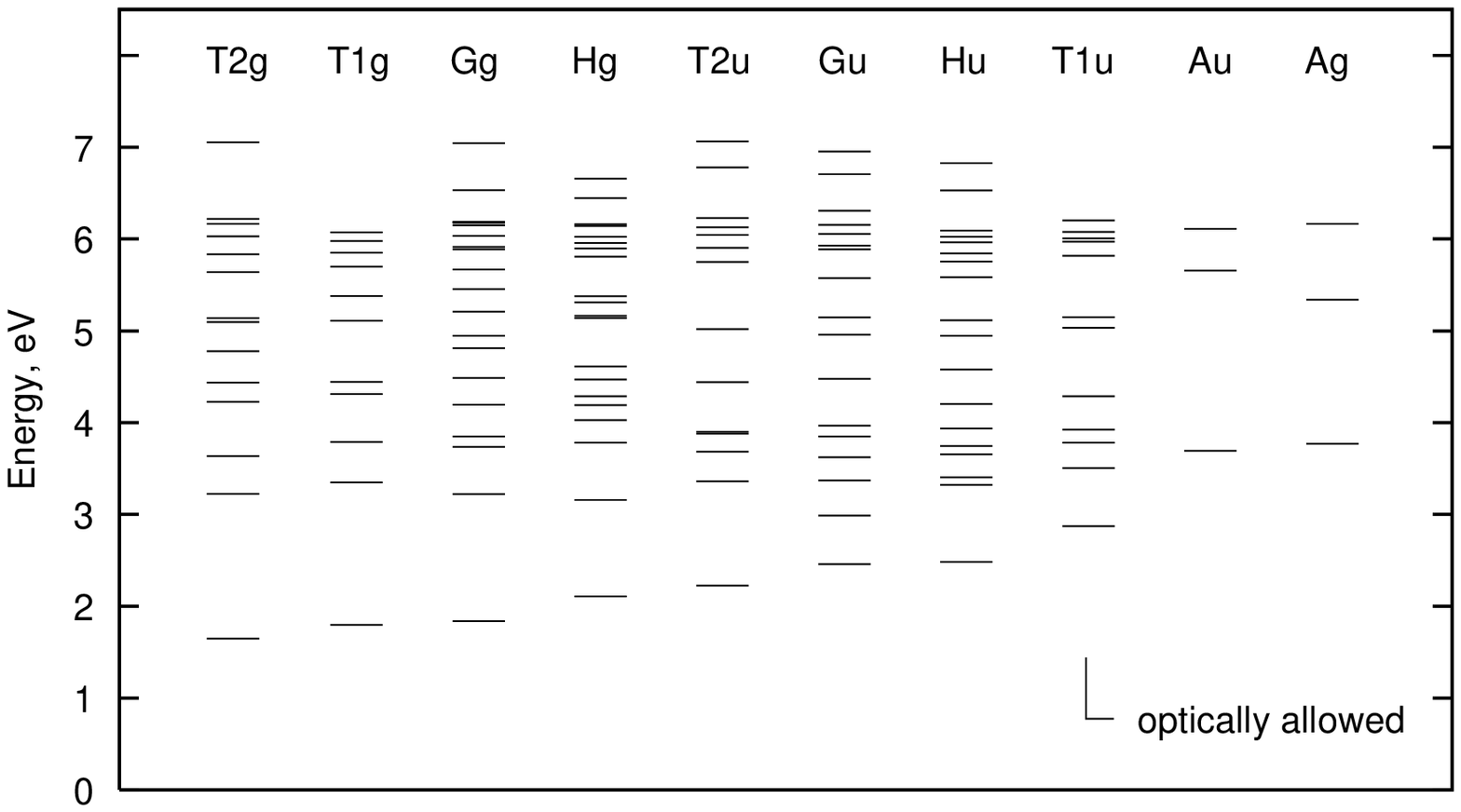,width=8cm}}
 {\small {\bf Fig.~2}. Excitation spectrum of C60 obtained using TDHF
with INDO/S semiempirical Hamiltonian parameterization.  Only $T_{1u}$
states have nonzero oscillator strengths.}
 \end{figure}

Fig.~2 shows the complete excitation spectrum obtained.  A total of
500 singlet excited states have been computed.  The excitation
energies were found to be degenerate 1, 3, 4, or 5 times in accordance
with multiplicities of irreducible representations of the I$_h$
symmetry group.  No symmetry-induced simplification of the problem has
been used.  A symmetry analysis was performed for each mode after it
was computed and the irreducible representation of the symmetry group
was assigned.

 \vskip 0.1 in
 {\small {\bf Table I}. C60 Experimental and theoretical electonic
excitation energies and experimental oscillator strengths.
Experimental values are from linear absorption in n-hexane
\protect\cite{koudoumas}.  Percent values are the deviations with
respect to the experiment.}
 \vskip 0.1 in
 \centerline{
\begin{tabular}{cccc}
\tableline
\tableline
\multicolumn{2}{c}{Absorption} &
RPA \\
\multicolumn{2}{c}{\ \ \ Experiment \protect\cite{koudoumas}\ \ \ } &
INDO/S \\
\multicolumn{2}{c}{ } &
\ \ (full space) \\
$\hbar\omega$, eV &
$f_{osc}$ &
$\hbar\omega$, eV \\
\tableline
     3.04 & 0.015 & 2.874\ \ (5\%) \\
     3.30 &       & 3.505\ \ (6\%) \\
     3.78 & 0.37  & 3.782\ \ (0\%) \\
     4.06 & 0.10  & 3.924\ \ (3\%) \\
     4.35 &       & 4.287\ \ (1\%) \\
     4.84 &       & 5.031\ \ (4\%) \\
     5.46 & 2.27  & 5.150\ \ (6\%) \\
     5.88 &       & 5.816\ \ (1\%) \\
          &       & 6.008          \\
          &       & 6.078          \\
     6.36 &       & 6.202\ \ (2\%) \\
\tableline
\tableline
\end{tabular}
}
\vskip 0.1 in

High symmetry of the molecule causes the majority of states to be
optically dark.  Only the states of $T_{1u}$ symmetry may have nonzero
oscillator strengths and show up in linear absorption \cite{C60book}.
It seems that the abundance of the singlet optically dark states below
the first optically allowed transitions is not fully realized.  The
present result may, therefore, shed some light onto a controversial
issue of an apparent anomalously fast singlet-to-triplet relaxation
\cite{C60book}.

The problem could have been simplified by taking symmetry
considerations into account {\em before} the RPA equations are solved.
It would be, however, opposite to the purpose of this letter, which is
to demonstrate in the first place the performance of the method for a
complex problem.  In particular, specific difficulties could have been
expected from the high level of degeneracies in the spectrum.  No
problems of that kind have been noticed.

In conclusion, a new method is proposed for solving RPA-type equations
with the computational effort comparable to that required to solve
static self-consistent field equations for the ground state.  The
method allows to compute low-energy excitonic states at the level of
theory which may be hard or impossible to achieve using conventional
techniques.

As suggested in Ref.~\onlinecite{ours_jcp}, calculation of the
electronic excitation energy at various nuclear configurations yields
effectively the excited-state adiabatic surface of the molecule,
provided that the ground-state adiabatic surface is known.  Thus, an
ability to compute the excitation energy at computational expense
comparable to the ground-state calculation can provide a long-sought
opportunity to perform realistic molecular-dynamics simulations of
photochemical reactions of large biological molecules.

{\bf Acknowledgements.}

I am grateful to I.L. Aleiner for discussions and for the opportunity
to pursue this research.

\end{multicols}
\end{document}